\documentclass[prl,twocolumn,showpacs]{revtex4}

\usepackage{times}
\usepackage{bm}
\usepackage[dvips]{graphics}
\usepackage{amsmath}
\usepackage{bm}
\usepackage{natbib}
\usepackage{color}
\usepackage{epstopdf}
\usepackage{graphicx}
\begin{document}

\title{Spin dynamics and relaxation in graphene dictated by electron-hole puddles}
\author{Dinh Van Tuan$^1$, Frank Ortmann,$^2$, Aron W. Cummings$^1$,  David Soriano$^1$ and Stephan  Roche,$^{1,3}$} 
 \affiliation{$^1$ICN2 - Institut Catala de Nanociencia i Nanotecnologia,
 Campus UAB, 08193 Bellaterra (Barcelona), Spain\\
$^2$ Institute for Materials Science, Dresden Center for Computational Materials Science, Technische Universit\"{a}t Dresden, 01062 Dresden, Germany  \\
 $^3$ICREA, Instituci\'{o} Catalana de Recerca i Estudis Avan\c{c}ats,
  08070 Barcelona, Spain}
\date{\today}
\begin{abstract}
The understanding of spin dynamics and relaxation mechanisms in clean graphene and the upper time and length scales on which spin devices can operate are prerequisites to realizing graphene spintronic technologies. Here we theoretically reveal the nature of fundamental spin relaxation mechanisms in clean graphene on different substrates with spin-orbit Rashba fields as low as a few tens of $\mu$eV.  Spin lifetimes ranging from 50 picoseconds up to several nanoseconds are found to be dictated by substrate-induced electron-hole characteristics. A crossover in the spin relaxation mechanism from a Dyakonov-Perel type for ${\rm SiO}_2$ substrates to a broadening-induced dephasing for ${\rm hBN}$ substrates is described. The energy dependence of spin lifetimes, their ratio for spins pointing out-of-plane and in-plane, and the scaling with disorder provide a global picture about spin dynamics and relaxation in ultraclean graphene in presence of electron-hole puddles.
\end{abstract} 

\pacs{72.80.Vp, 73.63.-b, 73.22.Pr, 72.15.Lh, 61.48.Gh} 
\maketitle


The tantalizing prospect of graphene spintronics was initiated by Tombros and coworkers \cite{Tombros2007}, who first reported long spin diffusion length in large area graphene. The small spin-orbit coupling (SOC) in carbon, plus the absence of a hyperfine interaction, suggested unprecedented spin lifetimes ($\tau_{s}$) at room temperature (from $\mu$s to ms) \cite{Hernando2006, Ertler2009, Gmitra2009, Ast2012,Ochoa2012, Han2014}. 

However, despite significant progress in improving graphene quality, resolving contact issues, and reducing substrate effects \cite{Tombros2007, Pi2010, Yang2007,Avsar2011, Zomer2012, Dlubak2012, Volmer2014, Guimaraes2014,Kamalakar2015}, the measured $\tau_{s}$ are orders of magnitude shorter, even for high-mobility samples. Extrinsic sources of SOC, including adatoms \cite{CastroNeto2009, Fedorov2013, Wojtaszek2013, Kochan2014} or lattice deformations \cite{Huertas-Hernando2009, Zhang2012}, have been proposed to explain this discrepancy. Moreover, the nature of the dominant spin relaxation mechanism in graphene is elusive and debated. The conventional Dyakonov-Perel (DP) \cite{DYA_SPSS13} and Elliot-Yafet (EY) \cite{YAF_SSP} mechanisms, usually describing semiconductors and disordered metals, remain inconclusive in graphene because neither effect can convincingly reproduce the observed scaling between $\tau_{s}$ and the momentum relaxation time $\tau_{p}$ \cite{Pi2010, Zomer2012}. Although generalizations of both mechanisms have been proposed, they do not allow an unambiguous interpretation of experiments \cite{Huertas-Hernando2009, Dora2010, Zhang2012,Ochoa2012,Roche2014a}.

It should be noted that the achieved room-temperature spin lifetime in graphene is already long enough for the exploration of spin-dependent phenomena such as the spin Hall effect \cite{Balakrishnan2013,Balakrishnan2015}, or to harness proximity effects as induced for instance by magnetic oxides \cite{Wang2015} or semiconducting tungsten disulphide \cite{Avsar2015}. However, a comprehensive picture of spin dynamics of massless Dirac fermions in presence of weak spin-orbit coupling fields is of paramount importance for further exploitation  and manipulation of spin, pseudospin and valley degrees of freedom \cite{Han2014,Roche2015,Son2006,Pesin2012}.  

In this study, we show numerically that a weak uniform Rashba SOC (tens of $\mu$eV), induced by an electric field or the substrate, yields spin lifetimes from 50 ps up to several nanoseconds. The dominant spin relaxation mechanism is shown to be dictated by long range potential fluctuations (electron-hole puddles) \cite{Adam2009}. For graphene on a ${\rm SiO}_{2}$ substrate, such disorder is strong enough to interrupt the spin precession driven by the uniform Rashba field, resulting in motional narrowing and the DP mechanism. We also find the ratio $\tau_s^{\bot}/\tau_s^{\parallel} \simeq 1/2$, demonstrating the anisotropy of the in-plane Rashba SOC field. For the case of a hexagonal boron nitride (hBN) substrate, the role of electron-hole puddles is reduced to an effective energy broadening and the spin lifetime is limited by pure dephasing \cite{Rashba2009,Dinh2014}. These situations, however, share a common fingerprint -- a M-shape energy dependence of $\tau_{s}$ that is minimal at the Dirac point. Taken together, our results provide deeper insight into the fundamentals of spin lifetimes in graphene dominated by electron-hole puddles.

\section*{Results}
\subsection{Disorder and Spin dynamics.} 
\begin{figure}[htbp]
\begin{center}
\leavevmode
\includegraphics[width=0.5\textwidth]{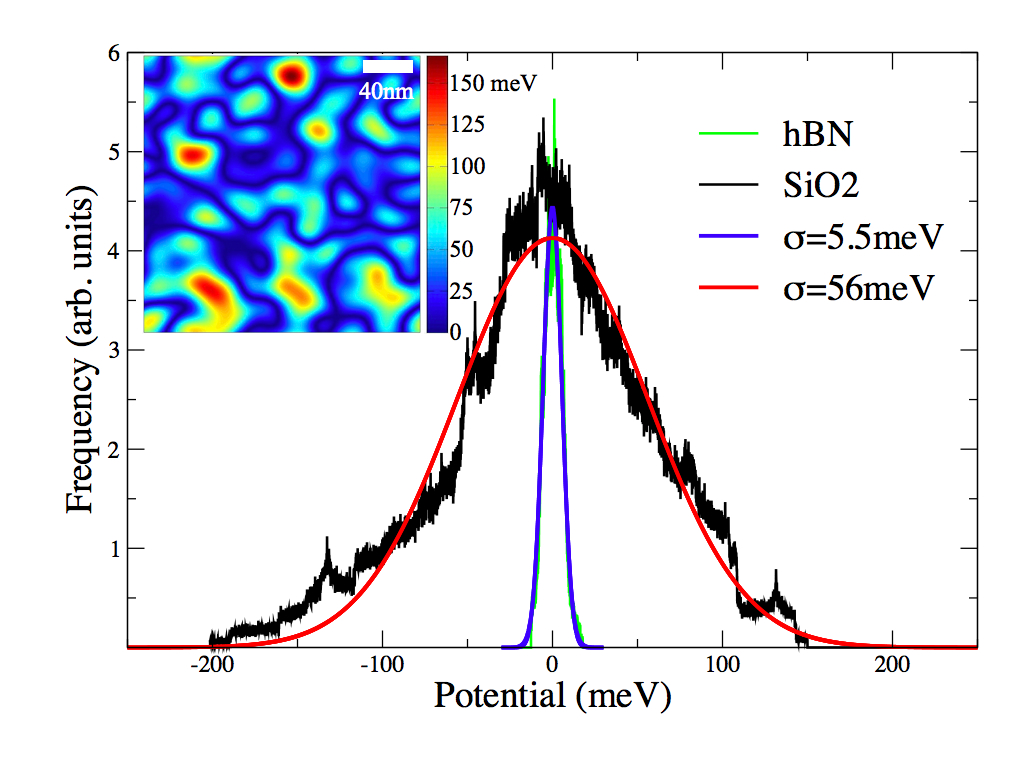}
\caption{Onsite energy distribution of the carbon atoms in the graphene sample, which mimics the chemical potential induced by hBN (green) and ${\rm SiO}_2$ (black) substrates together with their Gaussian fitting lines. Inset: Real space vizualization of the energy landscape for a graphene sample with $0.04\%$ Gaussian impurities (${\rm SiO}_2$ case).}    
\label{Fig1}
\end{center}
\end{figure}

Electron-hole puddles are real-space fluctuations of the chemical potential, induced by the underlying substrate, which locally shift the Dirac point \cite{Adam2009,Martin2008,Deshpande2009}. Since measured transport properties usually result from an average around the charge neutrality point, it is generally difficult to access the physics at the Dirac point. As shown by Adam and coworkers \cite{Adam2009}, electron-hole puddles can be modeled as a random distribution of long range scatterers, $V({\bf r})=\sum_{j=1}^{N}\epsilon_j\exp[-({\bf r}-{\bf R}_j)^2/(2\xi^2)]$, where $\xi=10$ and $30$ nm denote the effective puddle ranges for ${\rm SiO}_2$ and hBN substrates, respectively \cite{Martin2008, Zhangep}, and $\epsilon_j$ is randomly chosen within $\left[-\Delta,\Delta\right]$. Based on experimental data, typical impurity densities are $n_i=10^{12}$ ${\rm cm}^{-2}$ (0.04\%) for ${\rm SiO}_2$ and $n_i=10^{11}$ ${\rm cm}^{-2}$ (0.004\%) for hBN substrates \cite{Xue2011,Martin2008}. From such information, we can tune $\Delta$ to obtain suitable disorder profiles for the onsite energy of the $\pi$-orbital. Fig. \ref{Fig1} (main frame) shows the onsite energy distribution corresponding to hBN and ${\rm SiO}_2$ substrates, where the Gaussian profiles give standard deviations of $\sigma=5.5$ and $56$ meV, respectively. This allows us to extract $\Delta=50$ meV for ${\rm SiO}_2$ and $\Delta=5$ meV for hBN. The inset of Fig. \ref{Fig1} shows the energy landscape for a sample with 0.04\% Gaussian impurities (${\rm SiO}_2$ case). 

\begin{figure}
\begin{center}
\leavevmode
\includegraphics[width=0.5\textwidth]{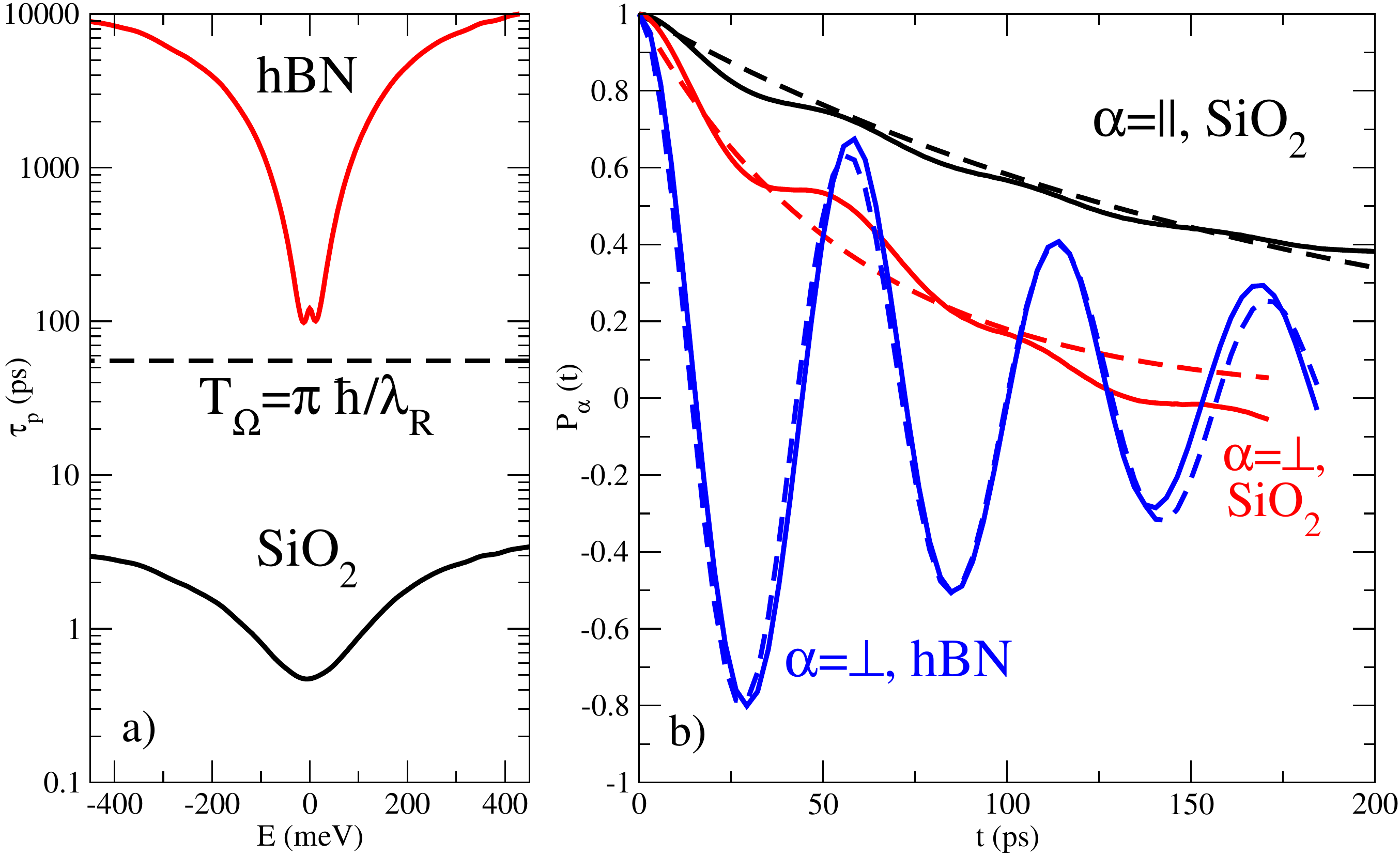}
\caption{ (a) Transport times for graphene on $\rm SiO_2$ and hBN substrates (solid black and red curves, respectively). The dashed line shows the spin precession time. (b) Time-dependent spin polarization for out-of-plane (solid red line) and in-plane (solid black line) spin injection for the ${\rm SiO}_2$ substrate, plus the fits to the exponential damping (dashed lines). The blue curves show the same information for the hBN substrate with out-of-plane injection.}    
\label{Fig2}
\end{center}
\end{figure}

To fully characterize the role of electron-hole puddles, we evaluate the transport time $\tau_p$ using a real-space order-N approach, which computes the diffusion coefficient $D(E,t)$. We extract $\tau_p$ from the saturation of $D(E,t)$ since $\tau_p=D_{max}(E)/2v_{F}^2(E)$ \cite{Roche2014b}. For numerical convenience, the calculations are made using $\Delta=0.1\gamma_0$ (with $\gamma_0$ the nearest neighbor hopping parameter), thus in absence of intervalley scattering \cite{Ortmann2011}, while the final values for hBN and ${\rm SiO}_2$ substrates are extrapolated from the numerical results using the scaling law

\begin{equation}
\tau_{p}\sim \frac{\sqrt{\pi n^*}\xi}{K_0}\frac{e^{\pi n^{*}\xi^2}}{I(\pi n^{*}\xi^2)},
\end{equation}

where $I_1(x)$ is the modified Bessel function of the first kind, $K_0=40.5n_i(\Delta/t)^2(\xi/\sqrt{3}a)^4$ is a dimensionless parameter dictating the strength of the Gaussian potential, and the carrier density $n^*$ is modified from the pristine graphene density $n$ by $n^{*}=|n|+\frac{K_0}{2\pi^{2}\xi^{2}}$ \cite{Adam2009,Rycerz2007,Klos2010}. The computed $\tau_p$ are shown in Fig. \ref{Fig2}(a) for both substrates. For $\rm SiO_2$, $\tau_p$ is on the order of a few ps, while for hBN $\tau_p$ is more than two orders of magnitude larger. The spin precession time, $T_\Omega = \pi\hbar/\lambda_R$, is shown for comparison.

\subsection{Spin dynamics and lifetimes in the presence of electron-hole puddles.} We now analyze the spin dynamics for puddles corresponding to the ${\rm SiO}_2$ and hBN substrates. The blue curve in Fig.2(b) shows the time-dependent spin polarization for the hBN substrate ($n_i=0.004\%$) at the Dirac point for an initial out-of-plane polarization, $P_\perp^{\rm hBN}(t)$ (see Methods). The polarization exhibits oscillations with period $T_\Omega = \pi\hbar/\lambda_{R}\simeq 55$ ps, corresponding to the spin precession induced by the Rashba field. Simultaneously, the polarization decays in time, and by fitting $P_\perp^{\rm hBN}(t) = \cos\left(2\pi t/T_\Omega\right) e^{-t/\tau_s}$, both $T_\Omega$ and the spin relaxation time $\tau_s$ can be evaluated.

Fig.2(b) also shows $P_\alpha^{\rm SiO_{2}}(t)$ for the ${\rm SiO}_2$ substrate ($n_{i}=0.04\%$) with initial spin polarization in-plane ($\alpha=\parallel$) and out-of-plane ($\alpha=\perp$). In contrast to the hBN case, for which $P_\perp^{\rm hBN}(t)$ exhibits significant precession, the disorder strength of electron-hole puddles for ${\rm SiO}_2$ is sufficient to interrupt spin precession. As a result, the polarization for ${\rm SiO}_2$ is better fit with $P_{\parallel/\perp}^{\rm SiO_2}(t)=e^{-t/\tau_s}$. The absence of precession for $P_\perp^{\rm SiO_2}(t)$ compared to $P_\perp^{\rm hBN}(t)$ is consistent with the ratio between transport time and precession frequency, since $\tau_{p}^{{\rm SiO}_{2}}/T_{\Omega}\ll 1 $ whereas $\tau_{p}^{{\rm hBN}}/T_{\Omega} > 1$.

To scrutinize the origin of the dominant relaxation mechanism, we first examine the spin lifetimes $\tau_s$ for the ${\rm SiO}_2$ case when rotating spin polarization (out-of-plane vs. in-plane), and varying the density of impurities ($0.04\%$, $0.08\%$, and $0.16\%$). Fig.3 shows the extracted $\tau_s$ in the out-of-plane (a) and in-plane (b) cases. The energy dependence of $\tau_s$ exhibits an M-shape increasing from a minimum at the Dirac point, with a saturation and downturn of $\tau_s$ for $E\geq 200$ meV. The values of $\tau_s$ range from 50 to 400 ps depending on the initial polarization and impurity density. We observe an increase of $\tau_s$ with $n_{i}$, which shows that a larger scattering strength reduces spin precession and dephasing, resulting in a longer spin lifetime, as described by the so-called motional narrowing effect \cite{Zutic2004}. Additionally, the ratio $\tau_s^{\bot}/\tau_s^{\parallel}$ (not shown) changes from 0.3 to 0.45 when $n_{i}$ is varied from $0.04\%$ to $0.16\%$. Such behavior is expected when enhanced scattering drives more randomization of the direction of the Rashba SOC field, which ultimately yields $\tau_s^{\bot}/\tau_s^{\parallel} = 0.5$ in the strong disorder limit \cite{Hernando2006,Ertler2009}. These results are fully consistent with the DP spin relaxation mechanism \cite{Huertas-Hernando2009,Zhang2012,Zutic2004}.

Fig.3(c) shows $\tau_s^\perp$ for the hBN substrate ($n_{i}=0.004\%$ and $0.016\%$) where a similar M-shape is observed. While $\tau_s^{\bot}(BN)$ is similar to $\tau_s^{\bot}({\rm  SiO_2})$ near the Dirac point, it is much larger at higher energies, reaching nearly 1 ns (for $\lambda_{R} = 37.4$ $\mu$eV). A striking difference is that the scaling of $\tau_s$ with $n_i$ is opposite to that of the $\rm SiO_2$ case, with an increase in puddle density resulting in a decrease in $\tau_s$, which indicates a different physical origin. For hBN, this behavior is reminiscent of the EY mechanism, but we will argue below that its origin is a different one. 

\begin{figure}
\begin{center}
\leavevmode
\includegraphics[width=0.5\textwidth]{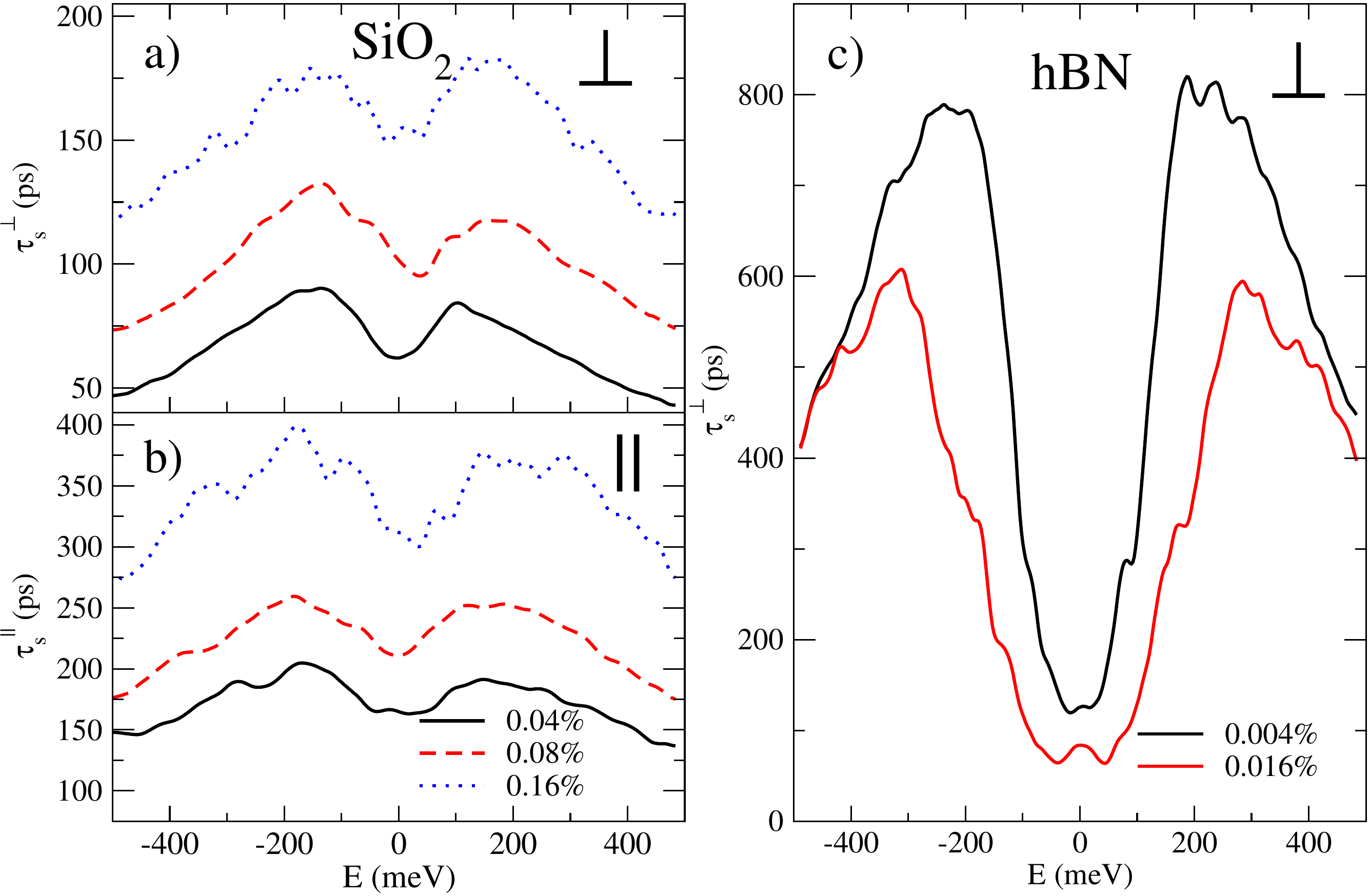}
\caption{Spin lifetimes for out-of-plane (a) and in-plane (b) spin injection for ${\rm SiO}_{2}$ substrate at impurity densities of 0.04\% (black solid curves), 0.08\% (red dashed curves), and 0.16\% (blue dotted curves ). (c)  Spin lifetime with out-of-plane spin injection for the hBN substrate at impurity densities of 0.004\% (black curve) and 0.016\% (red curve). }    
\label{Fig3}
\end{center}
\end{figure}

\subsection{Crossover in spin relaxation behavior for hBN and $SiO_2$ substrates.} Fig.4 provides a global view of our results, where we plot $\tau_s$ vs. $1/\tau_p$ for the $\rm SiO_2$ and hBN substrates (black and red symbols respectively) at the Dirac point and at $E = -200$ meV (closed and open symbols respectively). For low defect densities (hBN substrate), $\tau_s$ decreases strongly with decreasing $\tau_p$. However, with increasing defect density ($\rm SiO_2$ substrate) this trend reverses and $\tau_s$ scales almost linearly with $1/\tau_p$, according to the DP relationship $\tau_s = \nu \cdot (T_\Omega/2\pi)^2/\tau_p$. At $E = -200$ meV, $\nu = 1$, fitting the usual DP theory. At the Dirac point, the scaling is somewhat weaker, with $\nu = 1/4$. These results are reminiscent of those summarized in Fig. 5(a) of Drogeler {\it et al.} \cite{Volmer2014}, where spin lifetimes of graphene devices on $\rm SiO_2$ scaled inversely with the mobility, while devices on hBN appear to show the opposite trend.

While the $\rm SiO_2$ results of Fig.4 show DP behavior, the nature of the spin relaxation for weak electron-hole puddles is less clear. The fact that $\tau_s$ and $\tau_p$ decrease together suggests the EY mechanism, but we find $\tau_s \leq \tau_p$ near the Dirac point and $\tau_s \ll \tau_p$ at higher energies. This contrasts with the usual picture of EY relaxation, where charge carriers flip their spin when scattering off impurities, giving $\tau_s = \tau_p/\alpha$, where $\alpha \ll 1$ is the spin flip probability \cite{Ochoa2012}. Instead, this situation matches that described in Ref. \cite{Zutic2004}; when $\tau_p > T_\Omega$, the spin precesses freely until phase information is lost during a collision, in analogy to the collisional broadening of optical spectroscopy. More collisions result in a greater loss of phase, reducing $\tau_s$ with decreasing $\tau_p$. We verify this by removing the real-space disorder (setting $\Delta=0$) and modeling the electron-hole puddles with an effective Lorentzian energy broadening $\eta^*$. The results are shown in Fig.4 (main frame, blue dashed line), where we plot $\tau_s$ vs. $\eta^*$ at $E = -200$ meV (top axis). For small $\eta^*$, the scaling matches well with the real-space simulations of hBN, indicating that the puddles can be represented as a uniform energy broadening (See supplementary material). Larger values of $\eta^*$ lead to stronger mixing of different spin dynamics and $\tau_s$ saturates at very large $\eta^*$. There, the scaling of $\tau_s$ vs. $\eta^*$ clearly fails to replicate the DP behavior seen in the real-space simulations, since the effective broadening model does not induce the momentum scattering necessary for motional narrowing \cite{Zutic2004}.

\begin{figure}[htbp]
\begin{center}
\leavevmode
\includegraphics[width=0.5\textwidth]{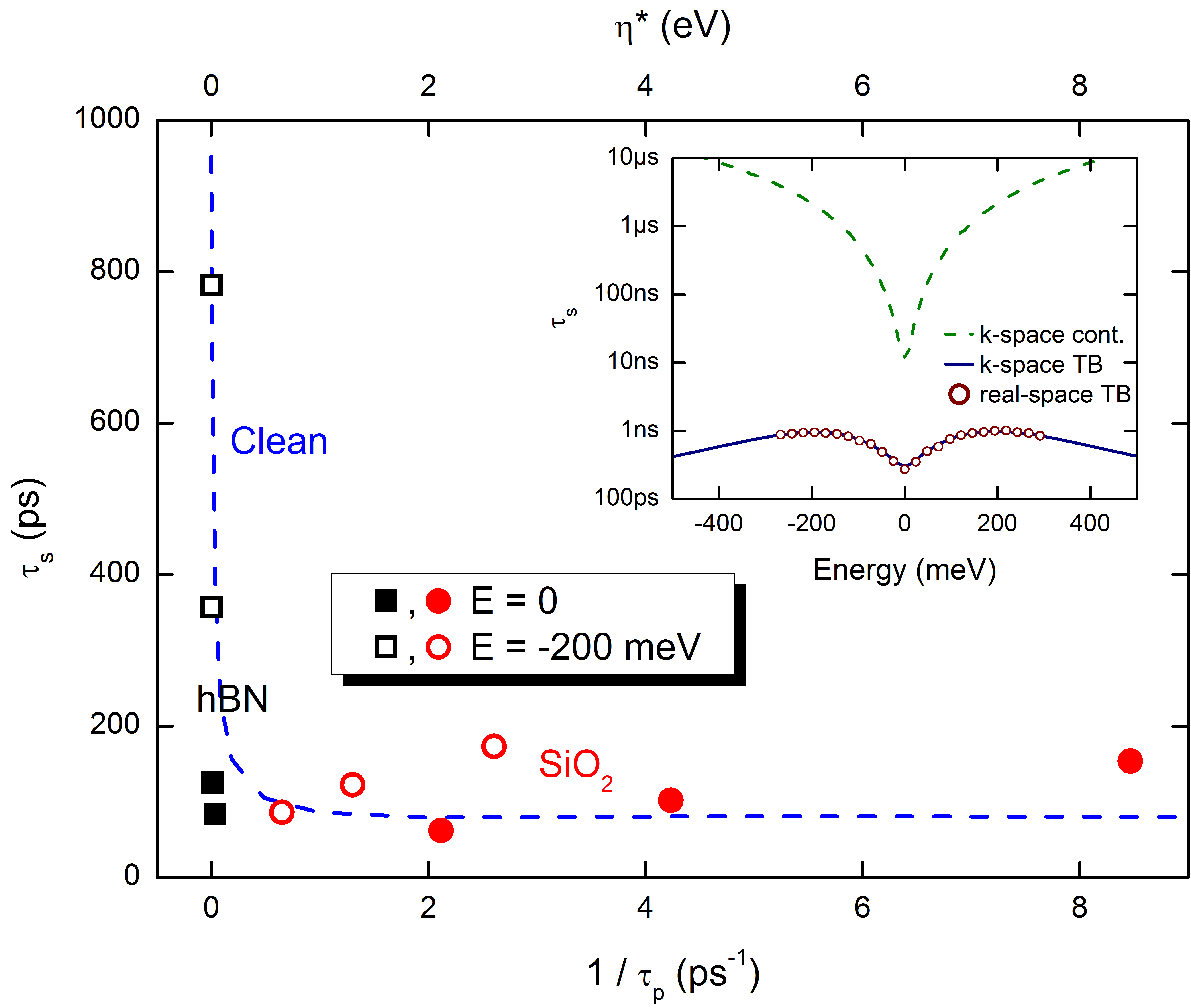}
\caption{Low-energy spin lifetimes versus $1/\tau_{p}$ (for initial out-of-plane spin polarization). Squares (circles) are for graphene on hBN (${\rm SiO}_{2}$) substrate. Closed (open) symbols are for spin relaxation at the Dirac point (at $E = -200$ meV). The blue dashed line shows the spin lifetime assuming only energy broadening (top axis). Inset: spin lifetime in absence of puddles computed using the TB model in real space (red circles) or $k$-space (blue solid line), and the low-energy model in $k$-space (green dashed line), with $\eta=13.5$ meV.}    
\label{Fig4}
\end{center}
\end{figure}

We finally explain the downturn of $\tau_s$ at the high energy wings of the M-shaped $\tau_s$ behavior in the hBN case. We compare the spin dynamics in the TB model (Eq. (\ref{TB}) in Methods) and the low-energy model in the absence of puddles ($\Delta=0$). In this regime $\tau_p \gg T_\Omega$, and spin dephasing and relaxation are driven by a combination of energy broadening and a nonuniform spin precession frequency. For the TB model, spin dynamics are calculated with the real-space approach and with a standard $k$-space approach and give identical $\tau_s$ (inset of Fig.4, red circles and blue solid line), indicating the equivalence of the real- and $k$-space approaches in the clean limit when accounting for the full TB Hamiltonian.  We observe that while for all models, the spin lifetime shows a minimum at the Dirac point (in agreement with experimental data, and explained by a strong spin-pseudospin coupling \cite{Rashba2009,Dinh2014}), spin transport simulations with the widely used low-energy Hamiltonian $\mathcal{H}^{(0)}$ (see Methods for $\mathcal{H}^{(0)}$ and green dashed line in Fig. 4 inset for results) clearly cannot capture the saturation and downturn of $\tau_s(E)$, i.e. its full M-shape. To qualitatively reproduce the M-shape of $\tau_s(E)$, the first-order term of the Rashba Hamiltonian, $\frac{\lambda_{R}a}{2}[k_{x} (\sigma_{x}s_{y}+\sigma_{y}s_{x}) + k_{y}(\sigma_{x}s_{x}-\sigma_{y}s_{y})]$, needs to be included in $\mathcal{H}^{(0)}$. This term introduces stronger dephasing at higher energy, driven by the anisotropy of the Rashba spin-orbit interaction \cite{Dinh2014}. 

In addition to their different energy dependence, the TB and low-energy models also yield very different spin lifetimes. A value of $\tau_s=10$ ns is obtained at the Dirac point for the low-energy model, which is two orders of magnitude larger than $\tau_s$ from the TB Hamiltonian, indicating a strong spin dephasing induced by the high-order $k$-terms. Interestingly, by studying the changes of $\tau_s(E)$ with respect to the Rashba SOC strength, we observe the scaling behavior $
\tau_s(E)\approx\beta(E)T_{\Omega}\approx \beta(E)\frac{\pi\hbar}{\lambda_R}$,  meaning the spin relaxes after a finite number of precession periods $\beta$ ($\beta \simeq 4.5$ close to the Dirac point), see Supplementary material. This suggests that dephasing is the limiting factor of spin lifetimes in the ultraclean case. We finally note that by taking $\lambda_R=5$ $\mu$eV (electric field of 1 V/nm \cite{Gmitra2009}), a spin lifetime of $\tau_s \simeq 1.4$ ns is deduced  at the Dirac point, whereas at higher energies $\tau_s$ could reach about 7 ns.

\section*{Discussion}
Our results show a clear transition between two different regimes of spin relaxation, mediated solely by the scattering strength of the electron-hole puddles. For hBN substrates, spin relaxation is dominated by dephasing arising from an effective energy broadening induced by the puddles, and $\tau_s$ scales with $\tau_p$. In contrast, for $\rm SiO_2$ substrates dephasing is limited by motional narrowing, leading to a DP regime with $\tau_s \propto 1/\tau_p$. Remarkably, both regimes exhibit similar values of $\tau_s$ at the Dirac point and a similar M-shape energy dependence (Fig.3), making it a signature of spin relaxation in graphene for all puddle strengths.  The crossover between both mechanisms occurs when $\tau_p\simeq T_{\Omega}$, which might have been realized in some experiments. This could explain some conflicting interpretations of experimental data  in terms of either Elliot-Yafet or Dyakonov-perel mechanisms \cite{Zomer2012}.

Our findings suggest alternative options for determining the spin relaxation mechanism in graphene from experimental measurements. Indeed, the typical approach, to examine how $\tau_p$ and $\tau_s$ scale with electron density and to assign either the EY or DP mechanism accordingly, is not always appropriate. For example, the EY mechanism in graphene is given by $\tau_s \propto E_F^2 \cdot \tau_p$, such that $\tau_s$ and $\tau_p$ would scale oppositely with respect to electron density if $\tau_p \propto 1/E_F$ \cite{Ochoa2012}. Similarly, for our results the scaling of $\tau_p$ and $\tau_s$ with energy suggest an EY mechanism near the Dirac point and a DP mechanism at higher energies, but Figs. 3 and 4 indicate a richer behavior. Therefore, to determine the spin relaxation mechanism it would be more appropriate to study how $\tau_s$ and $\tau_p$ scale with defect density or mobility at each value of the electron density.

Finally it should be noted that our simulations are performed using a constant Rashba spin-orbit coupling interaction, $\lambda_R$, which is different from the experimental situation where $\lambda_R$ will be increased at higher charge density owing to larger applied external electric field. This might explain why, especially for hBN substrate, the simulations show a larger variation of $\tau_{s}$ in energy than the gate voltage dependent spin lifetimes reported in the experiments \cite{Volmer2014,Guimaraes2014}.

\section{methods}
\subsection{Model of homogeneous SOC and electron-hole puddles.} The tight-binding (TB) Hamiltonian for describing spin dynamics in graphene is given by
\begin{eqnarray}
 \label{TB}
\mathcal{H} &=& -\gamma_0 \sum_{\langle ij \rangle} c_i^\dag c_j
        + i \frac{2}{\sqrt{3}} V_I \sum_{\langle\langle ij \rangle\rangle}
          c_i^\dag \vec{s} \cdot (\vec{d}_{kj} \times \vec{d}_{ik}) c_j \nonumber\\
        &+& iV_R \sum_{\langle ij \rangle}
          c_i^\dag \vec{z} \cdot (\vec{s} \times \vec{d}_{ij}) c_j,
\end{eqnarray}
where $\gamma_0$ is the nearest-neighbor $\pi$-orbital hopping, $V_I$ is the intrinsic SOC, and $V_R$ is the Rashba SOC. In the low-energy limit, this Hamiltonian is often approximated by a continuum model describing massless Dirac fermions, $\mathcal{H}^{(0)}=\hbar v_F \vec{\sigma}\cdot\vec{k}+\lambda_I \sigma_z s_z+\lambda_R \left( \vec{\sigma}\times\vec{s} \right)_z$, where $v_F$ is the Fermi velocity, $\hbar\vec{k}$ is the momentum, $\vec{s}(\vec{\sigma})$ are the spin (pseudospin) Pauli matrices, $\lambda_R=\frac{3}{2}V_R$, and $\lambda_I=3\sqrt{3}V_I$. The value $\lambda_I=12$ $\mu$eV is commonly used for the intrinsic SOC of graphene \cite{Gmitra2009} while the Rashba SOC is electric field-dependent. Here, we let $\lambda_R=37.4$ $\mu$eV, taken from an extended $sp$-band TB model for graphene under an electric field of a few V/nm \cite{Gmitra2009,Ast2012}. Higher-order SOC terms in the continuum model beyond $\mathcal{H}^{(0)}$ allow an extension to higher energy \cite{Rakyta2010}. 

\subsection{Spin dynamics methodology.} The time-dependent spin polarization of propagating wavepackets is computed through \cite{Dinh2014}

\begin{equation}
{\vec{P}}(E,t)=\frac{\langle\Psi(t)\rvert\vec{s}\delta(E-{\mathcal{H}})+\delta(E-{\mathcal{H}})\vec{s}~\rvert\Psi(t)\rangle}{2\langle\Psi(t)\rvert\delta(E-{\mathcal{H}})\rvert\Psi(t)\rangle}, \label{time-dependence0}
\end{equation}

where $\vec{s}$ are the Pauli spin matrices and $\delta(E-{\mathcal{H}})$ is the spectral measure operator. The wavepacket dynamics are obtained by solving the time-dependent Schr\"{o}dinger equation \cite{Roche2014b}, starting from a state $\rvert\Psi(t=0)\rangle$ which may have either out-of-plane ($z$-direction) or in-plane spin polarization. An energy broadening $\eta$ is introduced for expanding $\delta(E-{\mathcal{H}})$ through a continued fraction expansion of the Green's function \cite{Roche2014b}, and mimics an effective disorder. This method has been used to investigate spin relaxation in gold-decorated graphene \cite{Dinh2014}. Here, we focus on the expectation value of the spin $z$-component $P_{z}(E,t)= P_{\perp}(E,t)$ and the spin $x$-component $P_{x}(E,t)= P_{\parallel}(E,t)$.

\bibliography{biblatex-nature}


\section{addendum}
This work has received funding from the European Union Seventh Framework Programme under grant agreement 604391 Graphene Flagship. S.R. acknowledges the Spanish Ministry of Economy and Competitiveness for funding (MAT2012-33911), the  Secretaria de Universidades e Investigacion del Departamento de Economia y Conocimiento de la Generalidad de Catalu\~{n}a and the Severo Ochoa Program (MINECO SEV-2013-0295). F.O. would like to acknowledge the Deutsche Forschungsgemeinschaft (grant OR 349/1-1). Inspiring discussions with Sergio O. Valenzuela, Shaffique Adam, and Jaroslav Fabian are deeply acknowledged.

\end{document}